# Molecular Patterning and Directed Self-Assembly of Gold Nanoparticles on GaAs

Tianhan Liu[†], Timothy Keiper[†], Xiaolei Wang[‡], Guang Yang[§], Daniel Hallinan[§], Jianhua Zhao[‡], Peng Xiong[†]

[†] *Physics Department, Florida State University, Tallahassee, Florida 32306, USA*

[‡] *State Key Laboratory of Superlattices and Microstructures, Institute of Semiconductors, Chinese Academy of Sciences, Beijing, 100083, China*

[§] *Department of Chemical & Biomedical Engineering, Florida A&M University-Florida State University, Tallahassee, Florida 32310, USA*

*Email: pxiong@fsu.edu*

The ability to create micro/nano patterns of organic self-assembled monolayers (SAMs) on semiconductor surfaces is crucial for fundamental studies and applications in a number of emerging fields in nanoscience. Here, we demonstrate the patterning of thiol molecular SAMs on *oxide-free* GaAs surface by dip-pen nanolithography (DPN) and micro-contact printing (μCP), facilitated by a process of surface etching and passivation of the GaAs. A quantitative analysis on the molecular diffusion on GaAs was conducted by examining the writing of nanoscale dot and line patterns by DPN, which agrees well with surface diffusion models. The functionality of the patterned thiol molecules was demonstrated by directed self-assembly of gold nanoparticles (Au NPs) onto a template of 4-Aminothiophenol (ATP) SAM on GaAs. The highly selective assembly of the Au NPs was evidenced with atomic force microscopy (AFM) and scanning electron microscopy (SEM). The ability to precisely control the assembly of Au NPs on oxide-free semiconductor surfaces using molecular templates may lead to an efficient bottom-up method for the fabrication of nano-plasmonic structures.

Hybrid structures of functional molecules and solid-state (SS) materials have attracted extensive interest in surface nanoscience and molecular electronics.[1,2] Self-assembled monolayers (SAMs) are highly organized single layers formed spontaneously by organic molecules on a surface. The micro/nano patterning of organic SAMs on SS surfaces are common templates for bottom-up fabrication of hybrid devices.[3,4] Recently, a number of SAM patterning techniques, e.g., dip-pen nanolithography (DPN)[5–7] and micro-contact printing (μCP),[8,9] have been used to generate various micro/nano structures on SS surfaces.

Both DPN and μCP have been applied extensively for patterning thiol SAMs on gold surfaces. One of the applications of the SAMs is the directed assembly of nanostructures on a SS substrate, which has been realized on Au substrates by DPN[10,11] or μCP.[12,13] The nanostructure assemblies are typically facilitated by choosing proper combinations between the terminal group of the SAM and desired nanostructure, either through covalent bonding or electrostatic interactions. Among these nanostructures, NPs of Au and other noble metals as well as their variants have attracted great interest due to their unique and tunable electrical and optical properties, with applications in a variety of fields, such as electronics, optics, and chemical sensing. In particular, there has been intensive recent attention on the plasmonic behavior of arrays of Au NPs and their interactions with metallic, dielectric or semiconductor substrates.[14–17] Recently, the self-organization of Au NPs into ordered structures of arbitrary geometries on Au surfaces has been extensively demonstrated.[18–24] Extending directed self-assembly of Au NPs via molecular templates on semiconducting surfaces, especially those free of native oxides, is of both fundamental importance and practical utility. A prerequisite for this, however, is the micro/nano patterning of the molecular SAM on the surfaces of interest. To the best of our knowledge, there have been only a handful of studies of DPN and μCP on semiconductors, mostly on the native oxide surfaces. Ivanisevic and Mirkin first demonstrated DPN writing of hexamethyldisilazane (HMDS) on oxidized surfaces of Si and GaAs via the reaction of silazanes with oxide surfaces.[25] Wampler *et al.* succeeded in patterning of alkanethiols and short synthetic peptides by μCP and DPN on InP surfaces after removing the oxides with hydrofluoric acid.[26] A shortcoming of this method is the likelihood of the etched surface getting re-oxidized upon extended exposure to ambient air. Therefore, in order to produce high-quality SAM, particularly micro/nanoscale SAM patterns, on a pristine semiconductor surface, one needs to have a method of effective oxide removal and surface passivation. For many III-V semiconductors, such as GaAs and InAs, exposure to an ammonium polysulfide solution provides precisely these functions: It chemically removes the oxide layer and leaves a sulfur passivation layer which prevents reoxidation.[27–29] Compared to the etching methods with common acids for oxide removal, the ammonium polysulfide treatment is shown to protect the surface for days or even weeks.[30] Previously, we further demonstrated that by exposing a sulfur-passivated (Ga,Mn)As film to a 16-mercaptohexadecanoic acid (MHA) solution, the thiol terminals of the MHA molecules displace the sulfur and form SAMs on the (Ga,Mn)As. The MHA SAM is found to result in effective and controlled modification of the magnetic properties of the (Ga,Mn)As thin films.[30]

Here, we report the results of micro/nanoscale patterning of thiol SAM on oxide-free GaAs generated by DPN and μCP. With DPN, we not only designed and realized arbitrary patterns on GaAs, but also generated dots and lines of varying dimensions by controlling the deposition mode and time. The results provide the basis for a quantitative analysis of the diffusion of MHA molecules on GaAs. We also successfully generated large-scale MHA SAM micro-patterns on GaAs by μCP. Based on a μCP-generated ATP SAM template, we demonstrated a straightforward pathway for the directed self-assembly of Au NPs on pristine GaAs, as confirmed by SEM and AFM imaging.

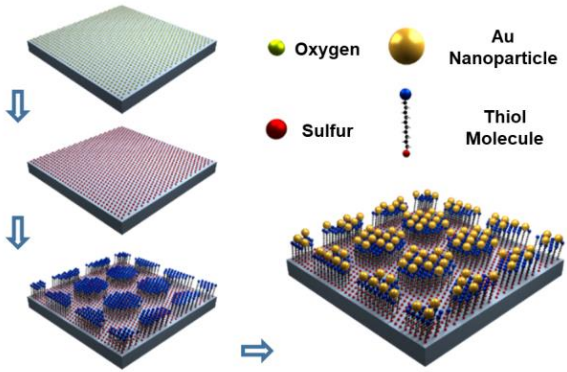

**Figure 1.** A schematic flow diagram of the fabrication procedure for the molecular template-directed self-assembly of Au NPs on GaAs, with key steps of (a) oxide removal and sulfur passivation formation by an ammonium polysulfide treatment, (b) formation of nano/micro scale thiol SAM patterns by DPN or μCP, and (c) self-assembly of Au NPs on the molecular template.

Figure 1 shows a schematic flow diagram of the general fabrication procedure. Initially, the GaAs (Both undoped wafers and Zn-doped wafers with carrier density varying from $5 \times 10^{18}$ to $5 \times 10^{19}$ $cm^{-3}$ were used.) is covered with a native oxide layer. An ammonium polysulfide $((NH_4)_2S_x)$ treatment selectively removes the oxide and results in a sulfur-passivated, oxide-free GaAs surface.[27–29] Next, thiol molecule SAMs are created directly on the sulfur-passivated GaAs surfaces by DPN or μCP. Finally, for directed self-assembly of Au NPs, SAMs of amine-terminated thiol molecules ATP are created and used for solution-based Au NPs assembly.

As a first step, we carried out a detailed investigation of SAM pattern formation by DPN and the diffusion of MHA molecules on the GaAs surface. The DPN was performed on sulfur-passivated GaAs. An AFM tip coated with MHA ink was brought into contact with the GaAs surface, and the tip was controlled by a software for the AFM to dictate the ink deposition on the substrate. Dots of varying sizes were obtained with a stationary tip by controlling the contact time and lines of varying widths by changing the speed of a moving tip. Immediately after a set of patterns were created, lateral force microscopy (LFM) images were taken, which discern regions of different SAMs and the GaAs based on the different frictional forces with the AFM tip. Figure 2 shows the LFM images of two sets of MHA SAM patterns of dots and lines. The darker regions in the figures correspond to the MHA SAM created by DPN, due to the smaller friction with the AFM tip than the GaAs substrate. To elucidate the kinetics of the MHA molecule diffusion associated with the DPN writing, the size of the dots and lines are analyzed with a diffusion model, and the fitting results are presented in Figure 2.

In dot writing, the AFM tip was stationary on the surface and molecules diffuse from the center outwards. With increasing deposition time, the dot size increases accordingly, as seen in Figure 2a. Quantitatively, the dot area exhibits a linear dependence on the deposition time, consistent with the prediction of the diffusion model based on 2D random walk,[31]

$$l^2 = 4D\Delta t, \quad (1)$$

where $l$ is the grid length, $D$ is the diffusion constant and $\Delta t$ is the time interval. The factor '4' arises from the two-dimensional random walk simulations, considering the four nearest neighbor sites, $(x+l, y)$, $(x-l, y)$, $(x, y+l)$ and $(x, y-l)$ available for a random jump from $(x, y)$.[31]

In our experiments, for a set deposition time $t$, the diameter of the resulting dot, $d$, can be determined by LFM. In Figure 2b, the diameters of the set of dots shown in Figure 2a are plotted against the square root of the deposition time; a linear relationship is evident. The data are consistent with the relation $r^2 = 4Dt$ or $d = 4\sqrt{Dt}$, where $r$ is the radius of the dot. However, a linear fitting of the data with $\sqrt{t}$ reveals a non-zero intercept, which can be attributed to the non-negligible size of the meniscus formed once the AFM tip was brought into contact with the substrate surface. The fitting in Figure 2b is in the form of $d = a\sqrt{t} + b$, yielding $a = 0.167 \pm 0.00586$ μm/s$^{1/2}$, $b = 0.336 \pm 0.0414$ μm. It implies a value of the diffusion constant $1.73 \times 10^{-11}$ cm$^2$/s, which is consistent with the diffusion coefficients reported in the literature.[25]

By DPN, line patterns were generated with a moving tip. In the process, the molecules diffuse from the footprint of the meniscus on the substrate surface and eventually become immobilized and form a SAM of finite width. Different line widths can be achieved by varying the velocity of the AFM tip. A number of publications have reported the relation between linewidth $w$ and tip velocity $V$.[32–36] Among them, a relatively accurate surface diffusion model was developed by Saha and Culpepper, taking into consideration the variation of the diffusion rate with the velocity of the AFM tip.[35] It describes mass conservation relation as

$$\rho V w = J + 2\rho V R, \quad (2)$$

where $\rho$ is the areal density of the ink molecules in the ordered SAM structure, $J$ is the ink diffusion rate, and $R$ is the radius of the meniscus footprint at the tip. Line writing by DPN is shown to exhibit increasing diffusion rate with increasing tip velocity in the model, which can be described by the relationship

$$V = \frac{2\pi D_s C_0}{\rho(w-2R)\ln(\frac{w}{2R})}, \quad (3)$$

where $D_s$ is the ink diffusivity and $C_0$ is the initial molecular concentration of the ink on the surface.

Figures 2c and 2d show the experimental results and the fitting to the diffusion model with parameters $R = 0.28$ μm and $\frac{D_s C_0}{\rho} = 6.4 \times 10^{-3}$ μm$^2$/s, respectively. The ink diffusivity extracted from the fitting results is consistent with the one fitted in the dot writing. The resulting sizes of the meniscus formations at the tip from the fittings in Figure 2b and 2d are similar, and much larger than the nominal 10 nm radius AFM tip, which is mainly due to the high ambient humidity of 55-60 %.

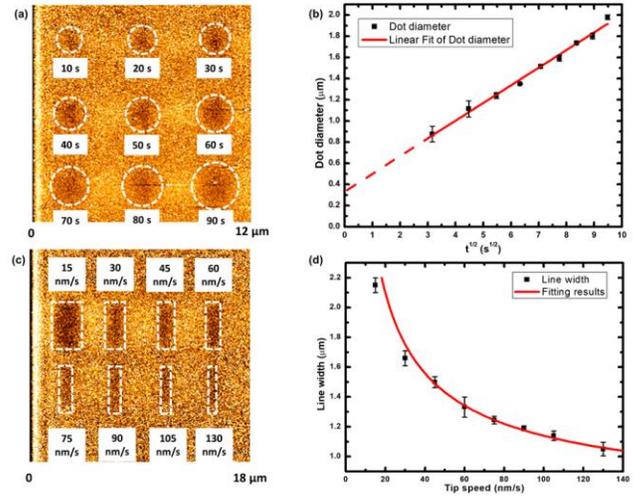

**Figure 2.** DPN dot and line writing and fitting results. (a) Dot patterns of MHA SAM on GaAs with deposition time from 10 s to 90 s, with 10 s increment. (b) Plot of the dot diameter vs. $\sqrt{t}$ and the fitting to a diffusion model. (c) Lines patterns of MHA SAM on GaAs drawn with tip speeds of 15, 30, 45, 60, 75, 90, 105 and 130 nm/s on GaAs substrate. (d) Plot of linewidth vs. writing speed and the fitting to a surface diffusion model.[35]

Based on the results and quantitative analyses of the DPN dot and line writing, we designed and fabricated a set of regular and arbitrary MHA SAM patterns on GaAs by DPN with respective optimal parameters. Figure 3a shows a cross of two single-pass lines drawn at tip speed of 100 nm/s, Figure 3b shows dots drawn with various deposition times, and Figure 3c is a set of squares drawn by multiple passes and the letters 'GaAs'. The latter demonstrates that with the proper surface preparation, regular and arbitrary patterns of thiol SAMs can be created on oxide-free GaAs in a highly predicable manner by DPN.

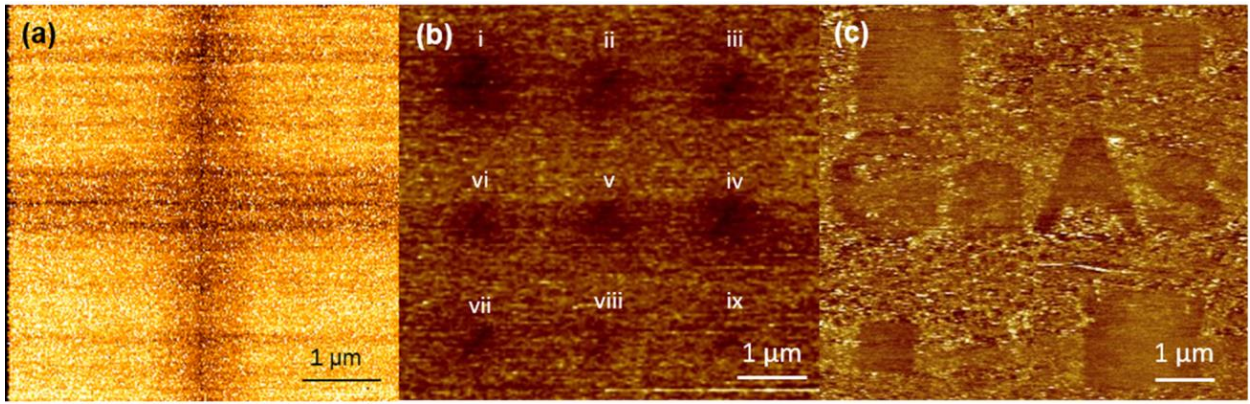

**Figure 3.** MHA SAMs features realized by DPN on GaAs, as imaged by LFM. (a) Crossed single-pass lines drawn with a tip speed of 100 nm/s. (b) An array of dots in diameters of (i) 1, (ii) 0.9, (iii) 0.8, (iv) 0.7, (v) 0.6, (vi) 0.5, (vii) 0.4, (viii) 0.3, and (ix) 0.2 μm. (c) Squares and letters showing the potential of generating arbitrary patterns of thiol SAM on oxide-free GaAs by DPN.

Another more rapid, large-area technique for generating microscale SAM patterns is μCP; we demonstrate that with the surface preparation, μCP can be used effectively to realize thiol SAM patterns on oxide-free GaAs. Here, we covered a PDMS stamp with 1 mM MHA ethanolic solution and then brought the stamp into contact with GaAs. Figure 4a shows schematically the procedure of transferring MHA from a PDMS stamps to oxide-free GaAs surface. The optical image of a PDMS stamp, where bright regions correspond to circular pillars, is presented in Figure 4b. Figures 4c-4e show LFM images of three samples fabricated under identical conditions with the exception of the μCP contact time. It is evident that with increasing contact time, there is increasing outward diffusion of the MHA molecules, resulting in larger circular patterns of the MHA SAM. This observation also provides strong evidence that the patterns are MHA SAM on GaAs, not artifacts of sulfur removal by the mechanical contact from the PDMS stamp.

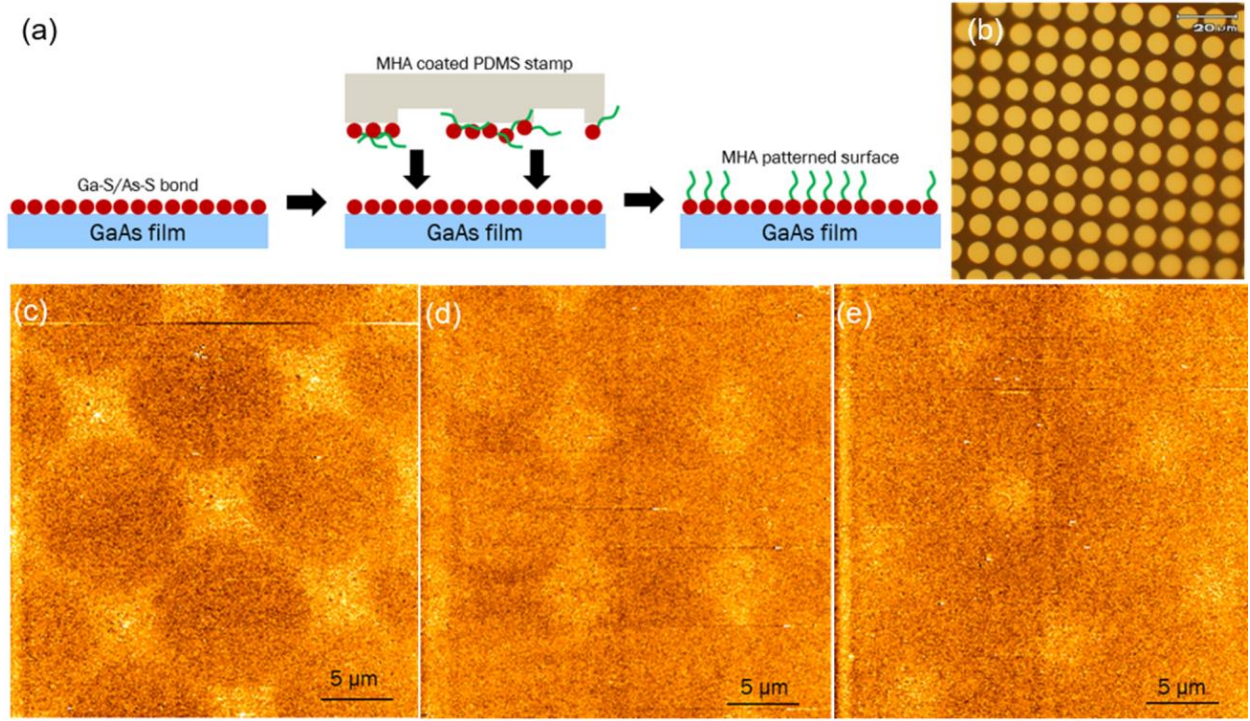

**Figure 4.** MHA SAM patterns realized by μCP on oxide-free GaAs. (a) Schematic of micro-contact printing. (b) Optical images of PDMS stamp. LFM images of μCP patterned MHA on GaAs with (c) 10 s, (d) 30 s, (e) 90 s of contact time.

Patterned molecular SAMs on semiconductor surfaces have a variety of potential applications. We demonstrate the functionality of the patterned molecules by using the thiol SAM patterns as a template for directed assembly of Au NPs on GaAs. We focused on the viability of the large-area thiol SAM patterns created by μCP on oxide-free GaAs. The Au NPs were synthesized by a modified Turkevich method (see details in the Experimental Section) and suspended in an aqueous solution. Figure S1 in the Supporting Information shows a TEM image of the Au NPs, from which the diameter of Au NPs is determined as 12.4 ± 0.5 nm, based on a Gaussian curve fit of the size distribution. For the Au NP assembly, we have compared the effectiveness of three different thiol molecules: MHA, ODT, and ATP with carboxylic, methyl, and amine terminal groups which are negatively charged, neutral, and positively charged, respectively. The Au NPs, which are negatively charged due to the surface citrate ions,[19] show weakest affinity to MHA and strongest affinity to ATP, as expected from the electrostatic interactions between the terminal groups and the Au NPs. Below we present the results of Au NP assembly on ATP templates. Figure 5a illustrates the fabrication procedure, which begins with the patterning of ATP on the sulfur-passivated GaAs by μCP, followed by the self-assembly of Au NPs. Figure 5b is an SEM image of the Au NP distribution on the GaAs; evidently, there is a concentrated distribution of Au NPs on the ATP regions inside the circle, with a distinct boundary with the outside region of few Au NPs. Figure 5c presents a close-up image of the area indicated by the red squared in Figure 5b. A quantitative analysis of two regions of the image by ImageJ (https://imagej.net/Welcome) yields Au NP areal densities in region 1 and 2 of 0.99 % and 9.2 % respectively, showing almost 10 times difference. Figure 5d shows an AFM topography image obtained by the tapping mode. The line scan in Figure 5e indicates that the height difference between the inside and outside of the circle is around 13 nm, further confirming the formation of monolayers of Au NPs in the ATP regions. The realization of directed assembly of Au NPs on semiconductors provides a potential pathway to rapid controlled fabrication of nanoplasmonic devices with the versatility afforded by the molecular recognition of the thiol SAMs for assembling arbitrary nanostructures.

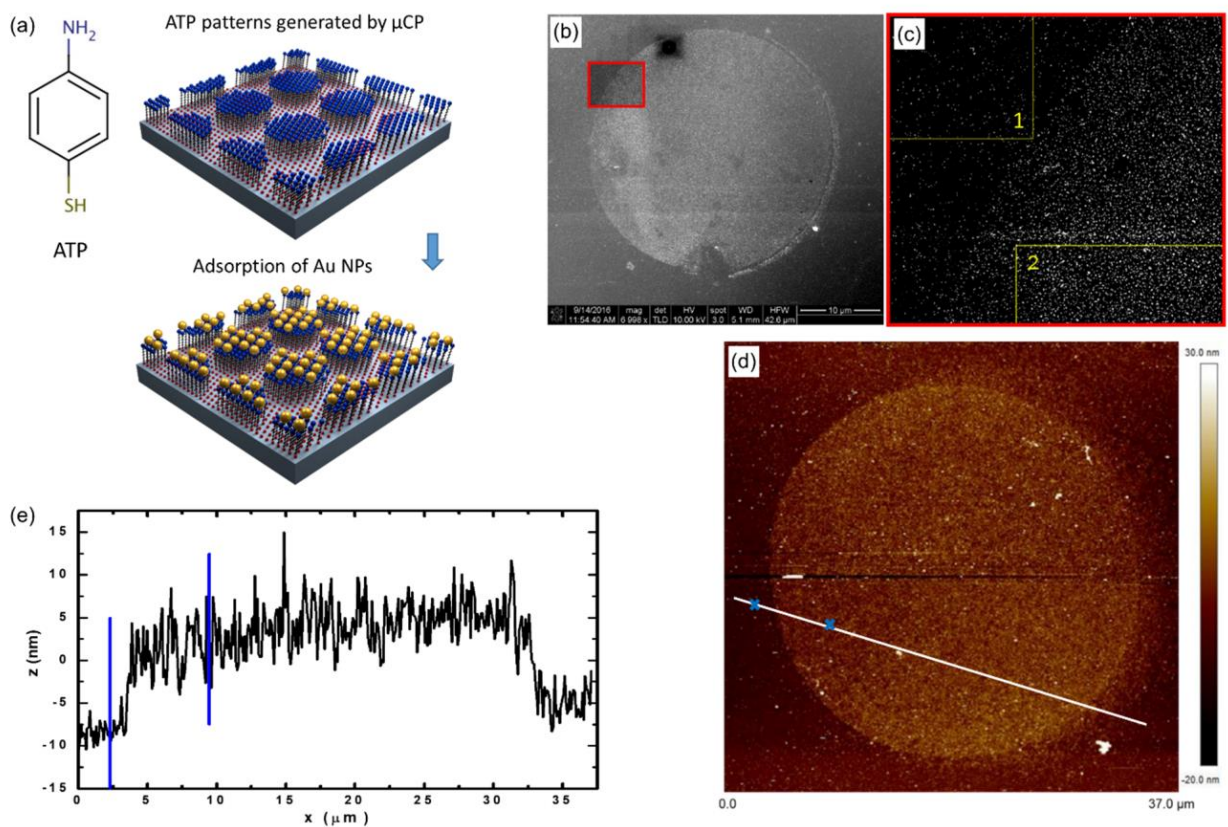

**Figure 5.** Directed self-assembly of Au NPs on ATP templates on GaAs. (a) Schematic illustration of the experimental procedures for Au NPs assembly on ATP templates on GaAs substrate. (b) SEM image of self-assembled Au NPs on GaAs surface. (c) Particle density analysis of the close-up SEM image in the red square in (b), with areal coverage 0.99% and 9.2% respectively for area 1 and 2 calculated by ImageJ. (d) AFM topography image of Au NPs and (e) the line scan showing a 13 nm height difference.

In summary, we have demonstrated a versatile approach to generating nano/microscale patterns of thiol molecular SAMs on oxide-free GaAs with arbitrary shapes and sizes by DPN and µCP. Both DPN and µCP show evidence of isotropic diffusion of thiol molecules on the GaAs surface. In particular, the results of DPN dot and line patterns were analyzed quantitatively and found to be well described by surface diffusion models. With these two techniques, we are capable of creating thiol SAMs with precise control over their size, geometry, and position on GaAs. The ability to controllably generate micro/nano patterns of organic SAMs on a pristine semiconductor surface holds great promise for molecular electronics and bottom-up fabrication of metal-semiconductor hybrid devices. The utility of the patterned SAMs was demonstrated by the directed self-assembly of Au NPs on an ATP SAM template on GaAs. The scheme can be readily adapted for the realization of nano/micro assemblies of NPs and nanocrystals on semiconductors, which may find applications in fields of optics, magnetics and nanoelectronics.

**Experimental Section**

*Materials and Sample Preparation:*

Both undoped and doped GaAs wafers were used. The undoped wafers were purchased from China Crystal Technologies Co., Ltd and the Zn-doped wafers with carrier density varying from $5 \times 10^{18}$ to $5 \times 10^{19}\ cm^{-3}$ were purchased from Wafer Technology LTD. The 21 % ammonium sulfide solution, MHA, ATP, Gold (III) chloride trihydrate (HAuCl$_4$ · 3H$_2$O, ⩾ 99.9% trace metals basis) and sodium citrate dihydrate (HOC(COONa) (CH$_2$COONa)$_2$ · 2H$_2$O ⩾ 99%) were purchased from Sigma-Aldrich.

GaAs samples (5 x 5 mm$^2$) were cleaned by sequential ultrasonication in acetone, methanol and isopropanol for 3 min and dried with nitrogen gas. Ammonium sulfide was diluted in deionized (DI) water at a volume ratio of 1:20. Then for each 100 mL solution, 0.5 g of elemental sulfur was added to make ammonium polysulfide solution ((NH$_4$)$_2$S$_x$). Subsequently, the GaAs samples were soaked in the (NH$_4$)$_2$S$_x$ solution for 10 min at 50 °C. Then, the samples were rinsed with DI water and ethanol to remove excess sulfur and blown dry with nitrogen gas. The passivation remains effective for days without significant re-oxidation of the surface based on our previous XPS results.[30]

Citrate-stabilized Au NPs were synthesized by a modified Turkevich method reported previously.[37] Briefly, aqueous solutions of $HAuCl_4$ (200 mL, 0.5 mM) and sodium citrate (10 mL, 38.8 mM) were separately brought to 100 °C. The latter was then rapidly injected to the former under vigorous stirring. A slow color change from light purple to ruby red was observed. The mixture was kept boiling for another 20 minutes until the color remained unchanged. Full reduction of the Au (III) to Au NPs is assumed due to the large stoichiometric excess of sodium citrate, which acts as a reductant.

*Fabrication Process:*

<u>Dip-Pen Nanolithography (DPN)</u>: An AFM tip (NanoInk Type A or Bruker's AFM probe MSCT-A) was cleaned with 10.15 W oxygen plasma (Expanded Plasma Cleaner from 'Harrick Plasma') for 10 min to increase the hydrophilicity. Then, it was immersed into 1 mM MHA ethanolic solution for about 30 s and the excess solvent was removed by nitrogen gas. During the writing process, the tip was operated in contact mode and moved to the desired region for drawing patterns controlled by the software in a NanoInk system. The ambient conditions were 23-25 °C and 55-60 % relative humidity for the characterization of DPN dot and line writing.

<u>Micro-contact Printing (μCP)</u>: The polydimethylsiloxane (PDMS) stamps were prepared by mixing Sylgard 184 silicone elastomer base and curing agent in a volume ratio of 10:1. Master molds with 5 μm-deep features based on silicon substrates patterned by photolithography were used as to cast the PDMS mixture. The PDMS was cured at 70 °C for two hours in an oven and carefully peeled from the master. Stamps with two different feature dimensions were used in the experiment: Stamp I has an array of 7 μm diameter circular pillars and 3 μm edge-to-edge separation, and stamp II has 30 μm diameter pillars and 10 μm separation. Prior to the SAM patterning, the stamps were sonicated in ethanol for 3 min. They were then covered by one drop of MHA solution, dried by nitrogen gas, and finally brought into contact with the substrate. After the stamping, samples were sonicated in ethanol and dried with nitrogen gas to remove superfluous molecules. The ambient conditions were 23-27 °C and 65-75 % relative humidity for all μCP experiments.

<u>Au NPs assembly</u>: The sulfur-passivated GaAs samples were patterned with ATP by μCP with stamp II, then left in an aqueous solution of Au NPs for 24 h. After the assembly, the samples were sonicated with deionized water and dried with nitrogen gas.

*Characterizations:*

DPN was performed with a commercial system from NanoInk, Inc. This system included a commercial AFM from Thermomicroscopes to allow observation by LFM. LFM images presented have a left-to-right scan direction. An Icon AFM from Bruker was used for AFM topography imaging. SEM images were obtained by a Nova NanoSEM from FEI with an accelerating voltage of 10.00 kV.

## ACKNOWLEDGMENT

We thank Steven Lenhert, David Van Winkle and Stephan von Molnár for helpful discussions.

The work at FSU is supported by NSF grant DMR-1308613. The work at IOS is supported by NSFC 11674312 and NSFC 11404323.

# Supporting Information

**Molecular Patterning and Directed Self-Assembly of Gold Nanoparticles on GaAs**

Tianhan Liu[†], Timothy Keiper[†], Xiaolei Wang[‡], Guang Yang[§], Daniel Hallinan[§], Jianhua Zhao[‡], Peng Xiong[†]

[†] *Physics Department, Florida State University, Tallahassee, Florida 32306, USA*

[‡] *State Key Laboratory of Superlattices and Microstructures, Institute of Semiconductors, Chinese Academy of Sciences, Beijing, 100083, China*

[§] *Department of Chemical & Biomedical Engineering, Florida A&M University-Florida State University, Tallahassee, Florida 32310, USA*

*Email: pxiong@fsu.edu*

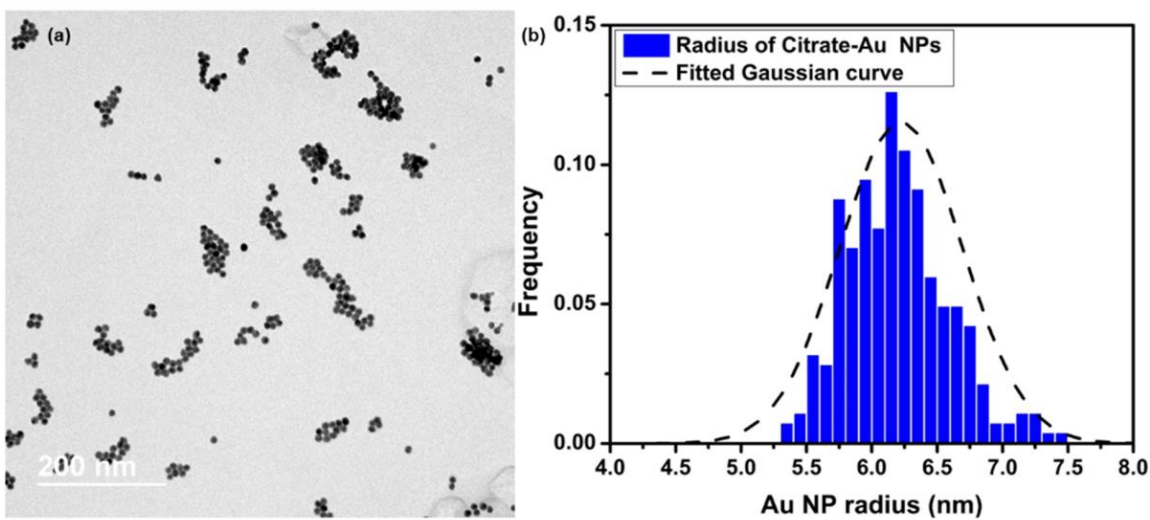

**Figure S1.** Au NPs and the size distribution. (a) TEM image of citrate Au NPs deposited on a carbon-coated TEM copper grid and (b) their size distribution histogram with fitted Gaussian distribution curve. The average size is 12.4 ± 0.5 nm based on the measurement of 500 Au NPs.[1]